\documentclass[amsmath,amssymb,superscriptaddress,showpacs,aps]{revtex4}

\usepackage{graphicx}
\usepackage{epsfig}
\usepackage{amssymb}
\usepackage{dcolumn}
\usepackage{bm}
\usepackage{subfigure}

\newcommand{\ba}{\begin{array}}
\newcommand{\ea}{\end{array}}
\def\br{\begin{eqnarray}}
\def\er{\end{eqnarray}}
\def\be{\begin{equation}}
\def\ee{\end{equation}}
\usepackage{hyperref}

\def\({\left(}
\def\){\right)}

\begin{document}

\title{QCD fixed points: Banks-Zaks scenario or dynamical gluon mass generation?}

\author{J. D. Gomez}
\email{john.gomez@ufabc.edu.br}

\affiliation{Universidade Federal do ABC, Centro de Ci\^encias Naturais e Humanas,
Rua Santa Ad\'elia, 166, 09210-170, Santo Andr\'e, SP, Brasil}

\author{A. A. Natale} 
\email{natale@ift.unesp.br}

\affiliation{Universidade Federal do ABC, Centro de Ci\^encias Naturais e Humanas,
Rua Santa Ad\'elia, 166, 09210-170, Santo Andr\'e, SP, Brasil}

\affiliation{Instituto de F\'{\i}sica Te\'orica, UNESP, Rua Dr. Bento T. Ferraz, 271, Bloco II, 01140-070, S\~ao Paulo, SP, Brazil}

\begin{abstract}
Fixed points in QCD can appear when the number of quark flavors ($N_f$) is increased above a certain critical value as proposed by Banks and Zaks (BZ). There is also the possibility
that QCD possess an effective charge indicating an infrared frozen coupling constant. In particular, an infrared frozen coupling associated to dynamical gluon mass generation (DGM)
does lead to a fixed point even for a small number of quarks. We compare the BZ and DGM mechanisms, their $\beta$ functions and fixed points, and within the approximations of this
work, which rely basically on extrapolations of the dynamical gluon masses at large $N_f$, we verify that between $N_f=8$ and $N_f = 12$ both cases exhibit fixed points at
similar coupling constant values ($g^*$). We argue that the states of minimum vacuum energy, as a function of the coupling constant up to $g^*$ and for several $N_f$ values, are
related to the dynamical gluon mass generation mechanism.
\end{abstract}


\maketitle

\section{Introduction}
It was observed in the well known work of Refs.\cite{gro,pol} that QCD is an asymptotically free theory if the number of quark flavors $N_f$ is smaller than a certain critical value.
When $N_f \leq 16$ the one-loop $\beta$-function is negative and the coupling constant diminishes as the energy is increased. Above this number the $\beta$-function becomes positive
indicating the increase of the coupling constant with the momentum. At two-loop order the $\beta$-function receives a contribution with a different signal as observed by Caswell
\cite{cas}, and although at high momentum this contribution is perturbatively small for a small number of flavors, its effect is not trivial if this number is increased, leading to
a zero of the $\beta$-function, and to the so-called Banks-Zaks (BZ) fixed point \cite{BZ}. The importance of a non-trivial perturbative fixed point is not only related to the interest on
the different QCD phases, but this fixed point may lead to a conformal window with possible interesting consequences for beyond standard model physics \cite{san}.

The $\beta$-function above two loops is scheme dependent, and the existence of the Banks-Zaks fixed point has been tested perturbatively 
for QCD up to four loops and in different schemes \cite{rit}. Therefore the fixed point location, as a function of $N_f$, has a small dependence  on the
renormalization scheme, in the order of perturbation theory that it was calculated, and, of course, the result can be considered reliable if
the next orders are indeed smaller. This motivated the interest of lattice simulations, not only for QCD, but also in different non-Abelian gauge
theories, and with different fermionic representations in order to determine the existence (or not) of this fixed point (see, for instance, \cite{app,pal,has} and references therein).
At four loops, as a function of $N_f$ and in the $\overline{MS}$ scheme, the QCD $\beta$-function for quarks in the fundamental representation is the following \cite{Beta4loop}:
\be
\beta(a_s)=-b_0 a_s^2-b_1 a_s^3-b_2 a_s^4- b_3 a_s^5 + {\cal{O}}(a_s^6) \, ,
\label{betaper}
\ee
where $a_s=\alpha_s/4\pi \equiv (g^2/4\pi)/4\pi$ and for SU(3) 
\br
b_0 & \approx &  11-0.66667 N_f \, ,\\
b_1 & \approx & 102 - 12.6667 N_f \, , \\
b_2 & \approx & 1428.50 - 279.611 N_f + 6.01852 N_f^2 \, ,  \\
b_3 & \approx & 29243.0 - 6946.30 N_f + 405.089 N_f^2 \, , \nonumber \\ 
    &    +    &   1.49931 N_f^3 \, .
\label{eqs:Coefficients}
\er
A zero of this $\beta$ function already appears if $N_f\geq 8$.

Parallel to the Banks-Zaks scenario there are other discussions about a possible infrared (IR) freezing (or IR fixed point) of the QCD coupling constant \cite{prosp,dok}.  
The concept of an IR finite QCD effective charge even for a small number of quarks was also extensively discussed by Grunberg \cite{gru}, and, in particular, he has also made an
interesting discussion about the relation of these effective charges with the conformal window originated through the Banks-Zaks expansion \cite{gru2}. These effective charges
naturally have a non-perturbative contribution that eliminates the Landau singularity for small $N_f$ \cite{gru2}. Another discussion following similar ideas can also be found in
Ref.\cite{brod}.   Other analysis about the transition region to these non-perturbative fixed points are presented in Ref.\cite{GiesAlkofer}.

There is one particular effective charge that generates a fixed point associated to the existence of a dynamically generated gluon mass \cite{nat3}. 
As shall be discussed ahead this charge is gauge invariant, has been obtained solving Schwinger-Dyson equations (SDE) for the QCD propagators and is consistent with lattice
simulations, but the main point that we shall discuss in this work, is concerned with a comparison between these different fixed points, i.e. the BZ and the one that appears when
gluons acquire a dynamically generated mass, which will be denominated DGM. Some of the questions that we will discuss here include: 1) Are the fixed points generated in these
different approaches numerically similar? 2) Are the anomalous dimensions associated to these fixed points varying in the same way and with similar values? 3) If these fixed points
are different for a given $N_f$ which one corresponds to a state of minimum energy?

The possible existence of a dynamically generated gluon mass was for the first time determined by Cornwall \cite{cor1}. 
Since this seminal work there were many others showing in detail that it is possible to write the SDE for the gluon propagator in a gauge invariant way \cite{pinch,papa3}, and from
it obtain a dynamical gluon mass \cite{s1,s2,s3,s4,s5,s6} and a gauge invariant IR finite frozen coupling \cite{g1,g2}. Lattice simulations are showing agreement with the SDE
\cite{la1,la2,la3,la4}. Even if the lattice simulations are performed in one specific gauge, it has been shown how the coupling constant obtained in the gauge invariant SDE approach
can be translated to a specific gauge, for instance the Landau gauge, and it was verified that the gap between the freezing values of the coupling constant obtained in different
gauges can be explained \cite{g1}. In Section II we will briefly discuss this effective charge and set the equations to show how it should vary with $N_f$.
This will allow us to compare the different $\beta$ functions, i.e. the BZ and the DGM ones. We will compute the different fixed points as a function of $N_f$. In Section III we will
check if these fixed points respect analyticity and compute their respective anomalous dimensions. In Section IV we provide some arguments about which scenario is the one leading
to the states of minimum of energy, or the one that should be chosen by Nature. In Section V we draw our conclusions. Within the approximations of this work, which rely basically
on extrapolations of the dynamical gluon masses at large $N_f$, we can compare the different $\beta$ functions (BZ and DGM) and their different fixed points, arguing that the  state
of minimum energy in QCD, as a function of the coupling constant and up to the critical value $g^*$, are related to the dynamical gluon mass generation mechanism only above $N_f=9$.

\section{Fixed points and dynamical gluon masses} 

Dynamical gauge boson masses appear in non-Abelian gauge theories as a consequence of the Schwinger mechanism, according to which if the gauge boson vacuum polarization
develops a pole at zero momentum transfer, the boson acquires a dynamical mass. The work of Ref.\cite{cor1} was the first one to obtain a gauge invariant SDE for the
gluon propagator, verifying the existence of the Schwinger mechanism in QCD. For this the introduction of the pinch technique was necessary, which combined with
the background field method, leads to a particular truncation of the SDE obeying abelian Slavnov-Taylor identities and to a gauge invariant gluon self-energy (${\hat{\Delta}}(k^2)$),
allowing us to build a renormalization group invariant quantity ${\hat{d}}(k^2)=g^2 {\hat{\Delta}}(k^2)$, whose solution can be written as
\be
{\hat{d}}^{-1}(k^2)= [k^2+m_g^2 (k^2)]\left\{ \beta_0 g^2 \ln \left(\frac{k^2+4m_g^2(k^2)}{\Lambda_{QCD}^2}\right)\right\} ,
\label{eqa1}
\ee
where $g$ is the coupling constant, $\beta_0=(11N-2N_f)/48\pi^2$, $N=3$, $m_g(k^2)$ the dynamical gluon mass
and $\Lambda_{QCD}$ is the characteristic QCD mass scale. The above equation can be
recognized as the equivalent at (large) momentum space to the inverse of the QCD Coulomb force, from where we can read the effective coupling constant 
\be
g^2(k^2)=\frac{1}{\beta_0 \ln\Big[\frac{k^2+4 \cdot m_{g}^{2}(k^2)}{\Lambda_{QCD}^2}\Big]}=4\pi\alpha_s (k^2),
\label{eq:Coupling}
\ee
which match asymptotically with the perturbative coupling constant.

Several aspects of this mechanism are illustrated in Refs.\cite{s12} and \cite{cor3}, respectively for the cases of QCD in four and
three dimensions. A recent review about dynamical gluon mass generation can be found in Ref.\cite{aguida}, which contains early references about this mechanism. Perhaps the many
technical details of the pinch technique and background field method hampered the dissemination of these results, but there is unequivocal evidence from large volume lattice
simulations that the gluon acquires a dynamical mass \cite{la3,la4}, the results are fully consistent with the SDE for the gluon propagator \cite{sx}, and consistent with many
phenomenological calculations that depend to some extent on the IR QCD behavior \cite{nat6}. 

The IR value of the dynamical gluon mass $m_g(k^2)$ is $m_g$ and it goes to zero at high momentum scales, with a running behavior roughly given by \cite{nat1}
\be
m_g^2(k^2) \approx \frac{m_{g}^{4}}{k^2+m_{g}^{2}}.
\label{eq:mg}
\ee
Without considering the running of the dynamical gluon mass with $k^2$, we have the following non-perturbative $\beta$-function \cite{cor2}
\be
\beta_{DGM}=-\beta_0 g^3\Big(1-\frac{4 m_g^2}{\Lambda^2}e^{-\frac{1}{\beta_0 g^2}}\Big),
\label{eq:betacornwall}
\ee
that will be denominated DGM $\beta$ function, where $\Lambda=\Lambda_{QCD}$.
Note that a more detailed expression for the non-perturbative coupling constant and dynamical mass can be found in \cite{g2}, however,
for simplicity, we will continue to use Eq.(\ref{eq:Coupling}) which reflects the gross behavior of this QCD IR finite coupling. One expression for the $\beta$-function taking into
account the running gluon mass as described in Eq.(\ref{eq:mg}) is given by \cite{nat2} 
\be
\beta_{DGM}(k^2)=-\beta_0 g^3 
\left(1-\frac{4 m_g^4 e^{-\frac{1}{g^2 \beta_0}}}{(m_g^2+k^2)\Lambda ^2}\left(1+\frac{k^2}{m_g^2+k^2}\right)\right),
\label{eq:betaMg_k2}
\ee
although this running barely affects the fixed point location that can be observed in Eq.(\ref{eq:betacornwall}). At this point we recall that the fixed point occurs at the
$k\rightarrow 0$ limit.  At this limit the fixed point of the non-perturbative $\beta$ function depends only on the $m_g/\Lambda$ ratio.

It can be demonstrated that the dynamical gluon mass generation imply in the existence of a non-trivial fixed point \cite{nat3}, and
more importantly: Recently we verified that the QCD coupling of Eq.(\ref{eq:Coupling}) freezes in the infrared limit, when only three quarks are operative, at one relatively small
value \cite{nat4}, and not too much different from the coupling values that would come out from the zeros of the perturbative $\beta$ function given by Eq.(\ref{betaper}) at some
large $N_f$! This fact certainly justifies the comparison of the different mechanisms (BZ or DGM). Moreover, this $\beta$ function may have an important consequence for the stability
of the standard model \cite{nat2}. Finally, to know how Eq.(\ref{eq:betacornwall}) or Eq.(\ref{eq:betaMg_k2}) vary with $N_f$ we must know how $m_g (k^2)$ varies with this quantity.
 
It has been observed in QCD lattice simulations that the dynamical gluon mass increases when $N_f$ increases \cite{aya}. Similar results
were observed in the solution of the Schwinger-Dyson equation (SDE) for the gluon propagator including dynamical quarks \cite{agui3}. We can
quote typical $m_g(0)$ values of $373, \,\, 427, \,\, 470$ MeV respectively for $N_f = 0, \,\, 2, \,\, 4$ quarks \cite{aya}. While several low energy phenomenological values,
obtained in the presence of two or three quarks, predict a somewhat larger value around $500$ MeV \cite{nat6}. Therefore we will conservatively assume that $373\leq m_g(0)\leq 500$
MeV for $N_f =0$. Note that it is not difficult to understand why $m_g$ should increase with $N_f$. When we increase the number of quarks we may expect that the strong force
diminishes (see, for instance, Eq.(\ref{eqa1})), because of the screening proportioned by extra quark loops. This force should be, at least asymptotically, proportional to the
product of the coupling constant times the gluon propagator. The coupling behaves as shown by Eq.(\ref{eq:Coupling}) and the propagator is roughly given by 
$\Delta (k^2) \propto 1/(k^2 + m_g^2(k^2))$. The only way to decrease the force as $N_f$ is increased is with larger $m_g(0)$ values.
Therefore, using the lattice data \cite{aya}, we will assume the following different scaling laws to describe the dynamical gluon mass evolution as a function of $N_f$:
\br
m_g^{-1}(N_f) & = &  m_{g_0}^{-1} \,\, (1-A_1 N_f)\label{eq:MgLinearPRD},\\
m_g^{-1}(N_f) & = &  m_{g_0}^{-1} \,\, e^{-A_2 N_f}\label{eq:MgExpPRD},
\er
where $m_{g_0}$ varies between $373$ and $500$ MeV and $A_1=0.05462$ and  $A_2=0.05942$. With the extreme $m_{g_0}$ values we obtain a band
of possible $\beta$ function curves, reflecting the uncertainty in our knowledge about the dynamical gluon mass.  Extrapolations like the ones of Eq.\eqref{eq:MgLinearPRD} and
\eqref{eq:MgExpPRD} have been used in Refs.\cite{aya,cap} and is probably the best that we can perform at the present status of lattice and SDE calculations. 

\begin{figure}[hbt]
\centerline{\includegraphics[width=8cm]{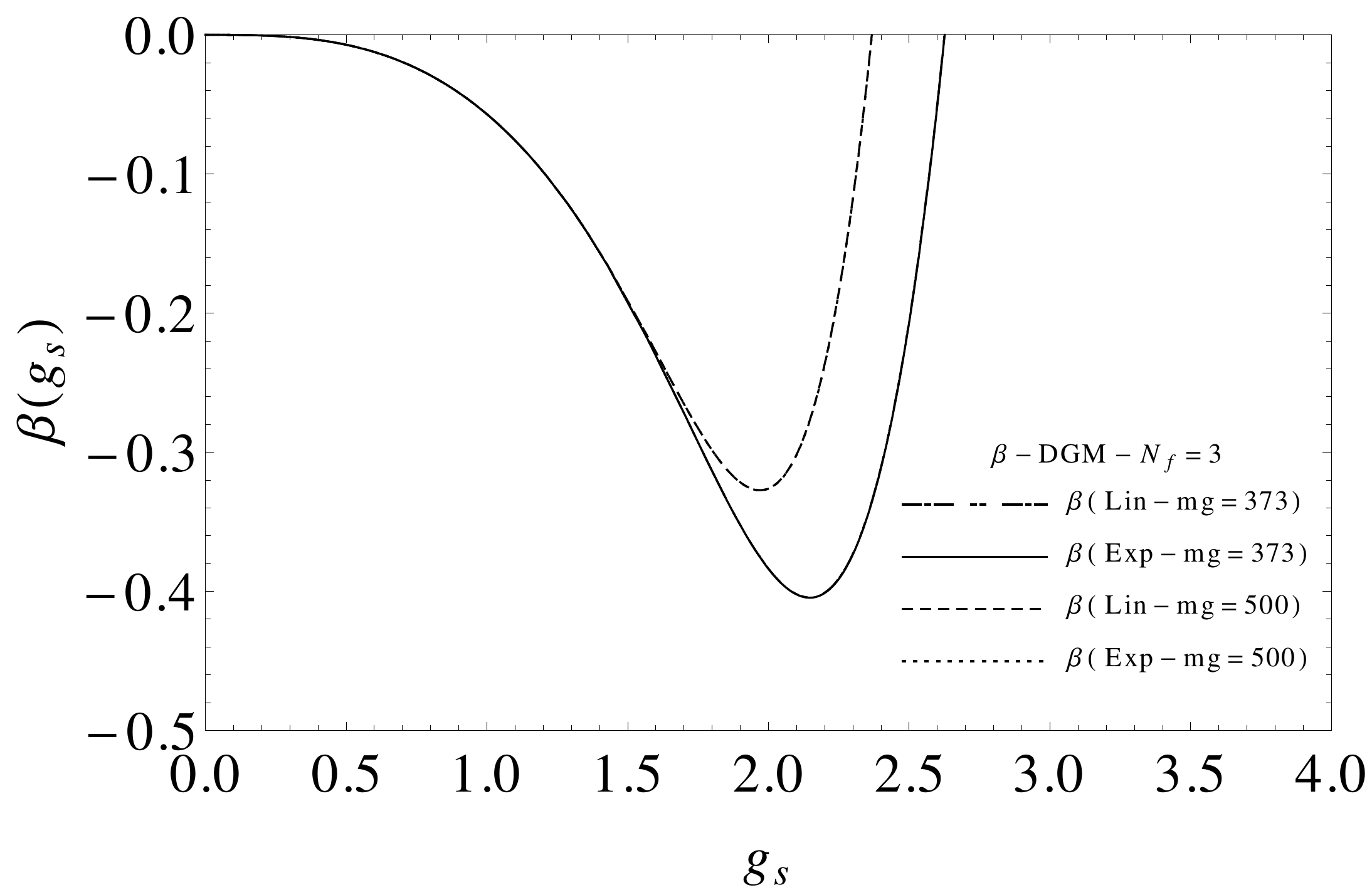}}
\caption{$\beta$ functions for the linear and exponential fits of Eqs.(\ref{eq:MgLinearPRD}-\ref{eq:MgExpPRD}) for $N_f=3$. \label{fig:BnpNf_3}}
\end{figure}

The different fits for the gluon mass evolution with $N_f$ are plotted in Fig.\eqref{fig:BnpNf_3}, where we can note that  with the 
same $m_{g_0}$,  there is no large difference between the linear and exponential fit for $N_f = 3$, but they start having a small difference above $N_f=6$.  The error between linear
and exponential fits for $N_f = 6, \, 8, \, 9,  \, 10 $ are $1.24\%, \, 3\%,\, 3.3\%, \, 5\%$ respectively. On the other hand we have also chosen an intermediate value of $m_{g_0}$
and computed the relative error with respect to $m_{g_0}=373$ MeV obtained with lattice data and the phenomenological value $m_{g_0}=500$ MeV.  With $m_{g_0}=440$ MeV we found an
error approximately of $5\%$ when comparing the different curves and with different (and large) $N_f$ values.
If we choose one appropriate fit we can compare the behavior of both $\beta$ functions BZ (Eq.\eqref{betaper}) and DGM with a running gluon mass (Eq.\eqref{eq:betaMg_k2}), comparing
different fixed points and discussing other consequences.


\begin{table}[htbp]
\caption{Fixed points ($\beta(g_s)=0$), i.e. critical coupling constant values ($g^*$), in the DGM approach with different 
values of $m_{g_0}$.}
  \begin{tabular}{@{}cccc@{}} \toprule
   $N_f\,\,\,$     &  $m_{g_0}$           &   $m_{g_0}$            &  $m_{g_0}$         \\
                   &      (373 MeV)       &       (440 MeV)        &   (500 MeV)        \\ \colrule
 3\hphantom{00}    & \hphantom{0} 2.63    & \hphantom{0} 2.47      &      2.37          \\
 6\hphantom{00}    & \hphantom{0} 2.79    & \hphantom{0} 2.64      &      2.54           \\
 8\hphantom{00}    & \hphantom{0} 2.98    & \hphantom{0} 2.83      &      2.74           \\
 9\hphantom{00}    & \hphantom{0} 3.11    & \hphantom{0} 2.97      &      2.87            \\
 10\hphantom{0}    & \hphantom{0} 3.29    & \hphantom{0} 3.14      &      3.03            \\ \botrule
 \end{tabular}
 \label{tbl:FP}
\end{table}

In Table \eqref{tbl:FP} we show the fixed points (coupling constant values) in the DGM approach (Eq.\eqref{eq:betaMg_k2}) with different number of flavors and three values of
$m_g(0)$.  It is possible to see that $g^*$ at the fixed point values become smaller when the $m_{g_0}$ is increased and also that for the same $m_{g_0}$ the value of the critical
coupling constant is increased when the number of flavors $N_f$ is increased. Observing these results and the small difference between the fits we will consider in all subsequent
calculations only the exponential fit of Eq.(\ref{eq:MgExpPRD}) and the $m_g(0)$ value of $440$ MeV. 

\begin{figure}[hbt]
\centering
 \subfigure[$\beta$ function of Eq.\eqref{eq:betaMg_k2}]
           {\resizebox{0.49\columnwidth}{!}{\includegraphics[width=0.4\textwidth,angle=0]{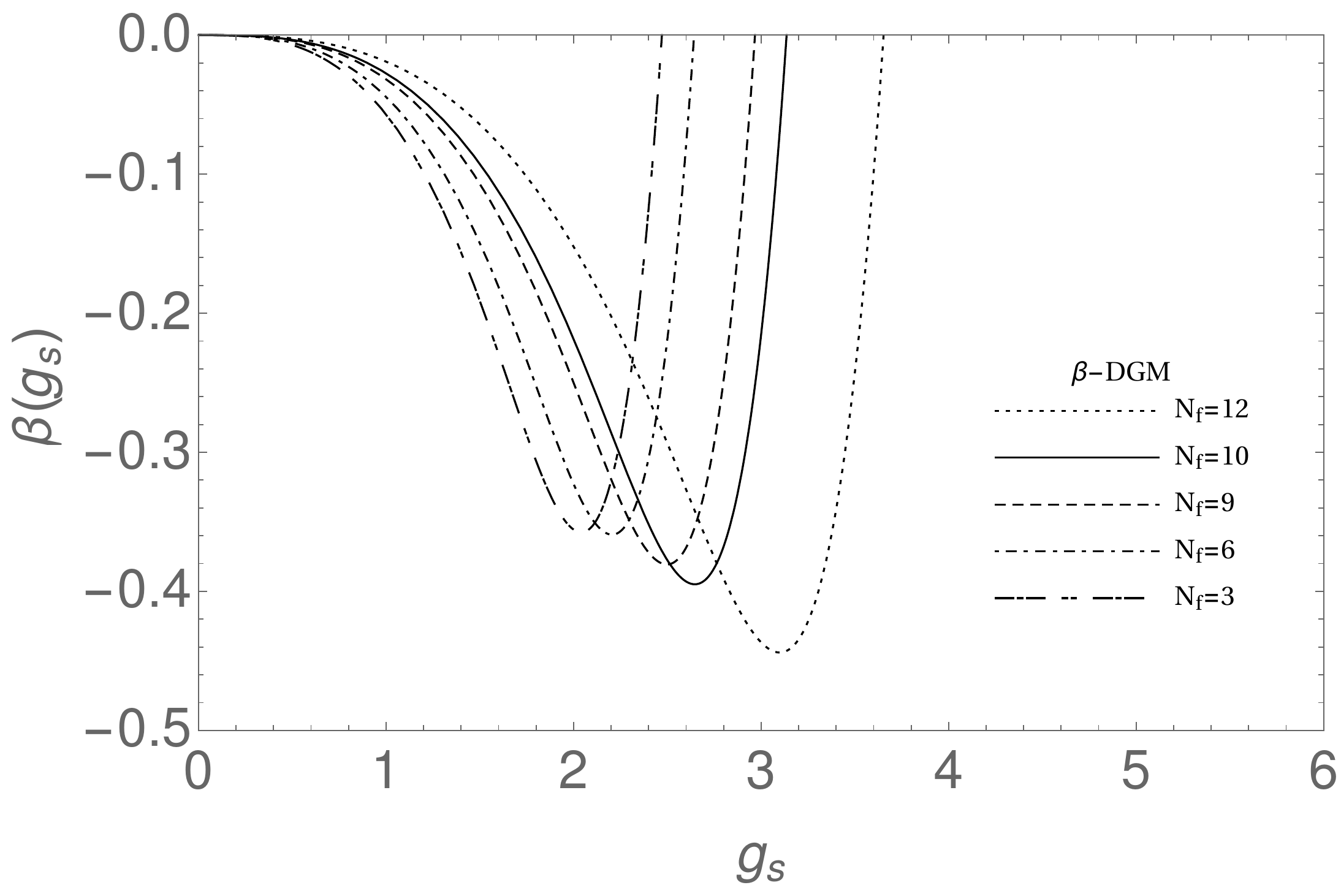}}}\hspace{0.05cm}
 \subfigure[$\beta$ function of Eq.\eqref{betaper}]
           {\resizebox{0.49\columnwidth}{!}{\includegraphics[width=0.4\textwidth,angle=0]{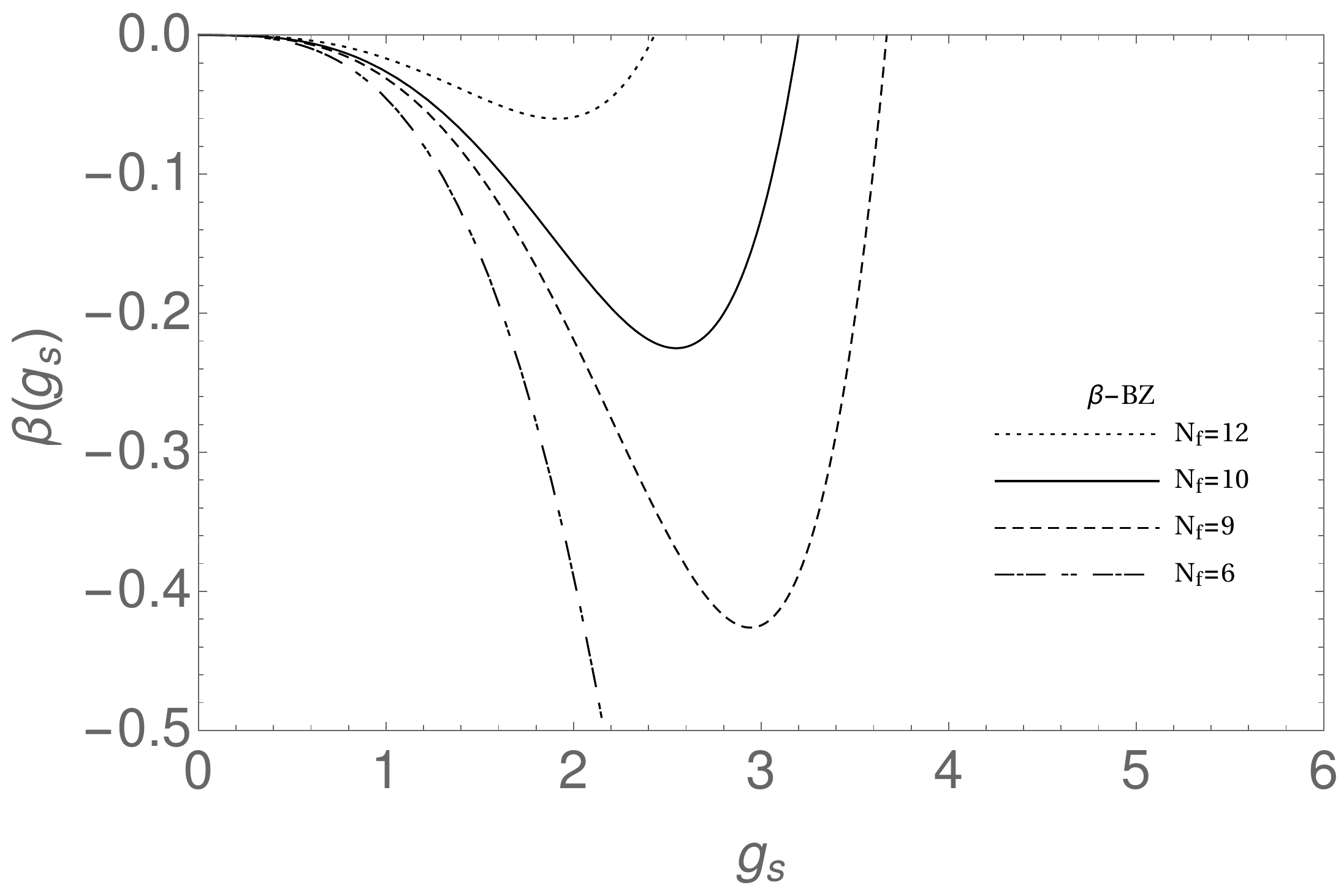}}}
 \caption{The DGM and BZ $\beta$ functions. Note that the non-trivial fixed points appear at approximately the same values of
the strong coupling ($g_s$), although they ``move" in different directions as $N_f$ is changed.}
 \label{fig:FP_Cor_Pert}      
\end{figure}

We show in Fig.\eqref{fig:FP_Cor_Pert} the $\beta$ functions for both cases (BZ and DGM). 
The fixed points in the BZ case start appearing for $N_f\geq 8$ [see Fig.\eqref{fig:FP_Cor_Pert}(b)], while in the DGM case they exist as long as
we have asymptotic freedom. The non-trivial fixed points appear at approximately the same values of the strong coupling ($g_s$). However as we increase
$N_f$ the BZ fixed points occur at smaller coupling constant values, while in the DGM we have exactly the opposite. The exact coupling
constant values for each fixed point can be better observed in Table \eqref{tbl:FP_BZ}.  It is interesting to comment how much the BZ fixed points are dependent on the order of
$\beta$ function calculated in perturbation theory. Although, two-, three-, and four-loop perturbative calculations of the $\beta$ function indicate an infrared fixed point in the
interval $8 \leq N_f\leq 16$, the recent five-loop calculation of this quantity \cite{Baikov} is showing that $N_f=8$ does not present a non-trivial perturbative fixed point at this
level (see, for instance, the discussion in Refs.\cite{Stevenson1}, \cite{Rittov1}). This higher loop result does not show the apparently mild convergent behavior of the lower loop
contributions and is currently being independently checked, therefore it is still premature to infer any possible behavior about
what is going to happen with the BZ scenario with higher order computations of the $\beta$ function.

\begin{table}[htbp]
 \caption{Values of the coupling constant ($g_s^*$) at the fixed points ($\beta(g_s^*)=0$) for both approaches (BZ and DGM), with $N_f$ between 6-13.}
  \begin{tabular}{@{}ccc@{}}\toprule
   $N_f$ \hphantom{00}    & \hphantom{00}       BZ       & \hphantom{0000}    DGM     \\ \colrule
     6 \hphantom{00}      & \hphantom{00}     $\ast$     & \hphantom{0000}    2.64     \\
     7 \hphantom{00}      & \hphantom{00}     $\ast$     & \hphantom{0000}    2.73     \\
     8 \hphantom{00}      & \hphantom{00}      4.41      & \hphantom{0000}    2.83     \\
    10 \hphantom{00}      & \hphantom{00}      3.20      & \hphantom{0000}    3.13     \\
    11 \hphantom{00}      & \hphantom{00}      2.80      & \hphantom{0000}    3.36     \\
    12 \hphantom{00}      & \hphantom{00}      2.43      & \hphantom{0000}    3.65     \\
    13 \hphantom{00}      & \hphantom{00}      2.06      & \hphantom{0000}    4.08     \\ \botrule
   \end{tabular}
   \label{tbl:FP_BZ}
\end{table}

Recall that we are comparing quantities obtained in different schemes. The non-perturbative effective coupling constant given by 
Eq.(\ref{eq:Coupling}) has been determined as a function of Green's functions obtained from SDE solutions, through the combination of the pinch technique with the background field
method. Such approach is gauge and renormalization group invariant, i.e. they are independent of any renormalization mass $\mu$ \cite{cor4,aguida}.  Note that at the end the fixed
point is only a function of $m_g/\Lambda$. This means that in principle we may have certain stability in the fixed point determination. However the SDE for the gluon propagator, from
where it is obtained part of the information leading to the infrared coupling, has to be solved imposing that the non-perturbative propagator is equal to the perturbative one at some
high-energy scale $(\mu )$, or comparing the SDE propagator to the lattice data. After this a $\mu$ independent coupling is obtained through one specific relation of two point
correlators. On the other hand we may say that the BZ $\beta$ function has a scheme dependence (${\overline{MS}}$) above the two-loop level, but its fixed point is relatively stable
\cite{rit}, and we may also say that the comparison between the different scenarios that we have been discussing is worthwhile. Finally, the DGM coupling moves to higher $g_s$ values
as we increase $N_f$, and at some moment even the non-perturbative method used to obtain this quantity may fail. Therefore in the next section we will verify if these different
$\beta$ functions are analytic up to the fixed point and what can be said about their anomalous dimension.

\section{Analyticity and anomalous dimension}
The renormalization group behavior of the coupling constant is constrained by the analyticity condition as proposed by Krasnikov \cite{Krasnikov} as
\be
\bigg\lvert \alpha_s \frac{d}{d\alpha_s}\Big(\frac{\beta(\alpha_s)}{\alpha_s}\Big) \bigg\rvert \leq 1.
\label{eq:Analyt}
\ee
This is a perfect condition to test if the different fixed points, or critical coupling constants, discussed in the previous sections can be considered still small enough to be
reliable, even if they were obtained with a non-perturbative method as in the DGM case. The problem of using Eq.(\ref{eq:Analyt}) is that it was obtained in the so called ``natural"
scheme, where the ratio $\beta (\alpha_s)/\alpha_s$ is modified, as we change from one scheme to the other, only by a multiplicative constant, which is not a significative change
near a fixed point. Nevertheless we have a condition constraining the coupling constant renormalization group behavior in one specific scheme, the BZ coupling determined up to four
loops in the $\overline{MS}$ scheme, and one non-perturbative coupling that is renormalization group invariant but obtained in one truncation dependent scheme. Many discussions on
the conformal window seems to indicate some stability in the critical coupling constant values obtained in different schemes \cite{rit,gra}. This fact does not justify a fully formal
comparison of these different quantities, although, as we shall see, we still can learn from it as well as extract some valuable information. Of course, a complete solution of all
these scheme differences, i.e. obtaining Eq.(\ref{eq:Analyt}), the BZ and DGM $\beta$ functions in one scheme independent way, is out of the scope of this work.

\begin{figure}[htb]
\centering
 \subfigure[Analyticity condition applied to the DGM $\beta$ function for different $N_f$. For each $N_f$ the inequality \eqref{eq:Analyt} is satisfied only below one specific
           $\alpha_s$ (or $g_s$) value.]
           {\resizebox{0.49\columnwidth}{!}{\includegraphics[width=0.4\textwidth,angle=0]{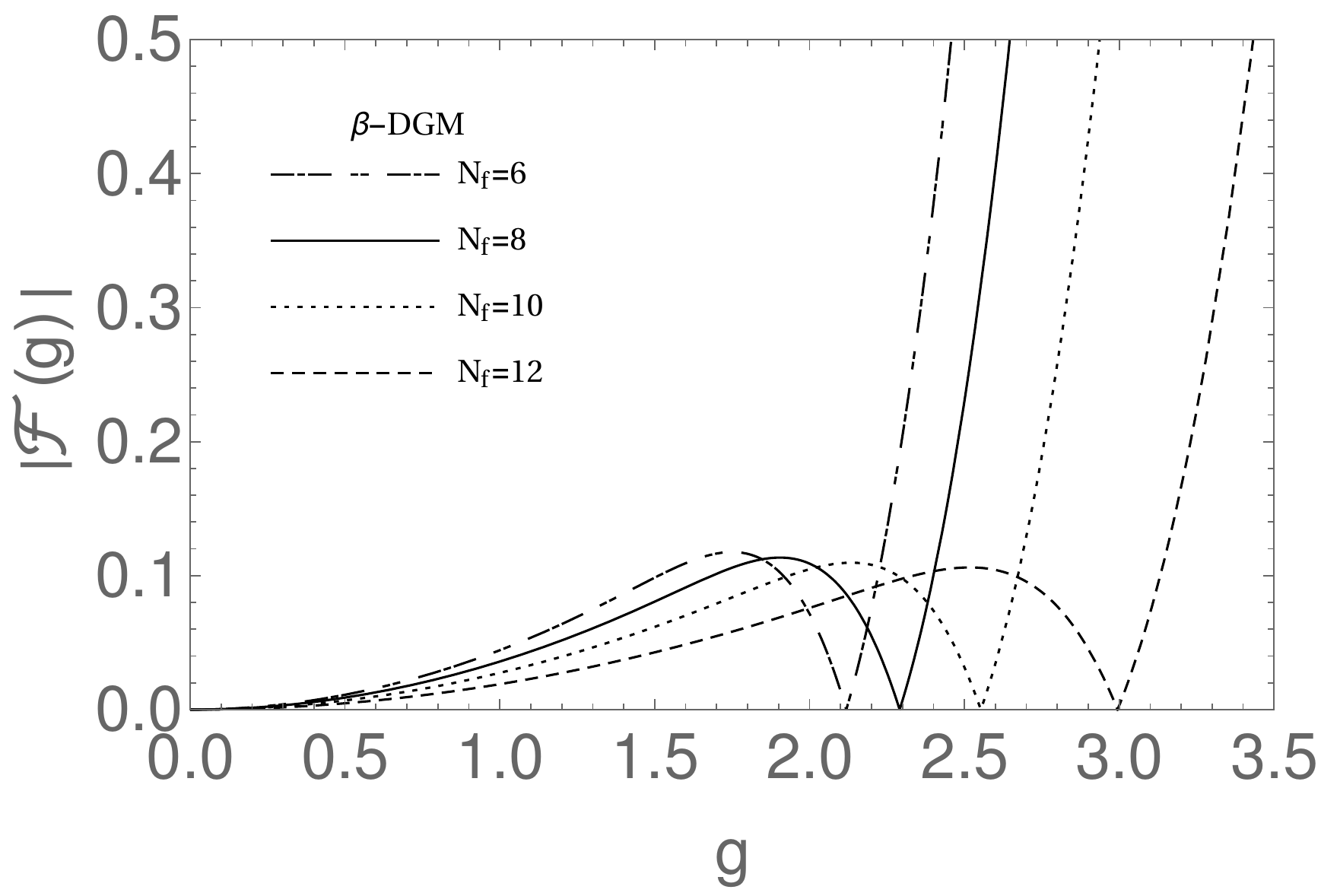}}}\label{fig:Cornw_Anali}\hspace{0.05cm}
 \subfigure[Analyticity condition applied to the perturbative (BZ) $\beta$ function for different $N_f$. ]
           {\resizebox{0.49\columnwidth}{!}{\includegraphics[width=0.4\textwidth,angle=0]{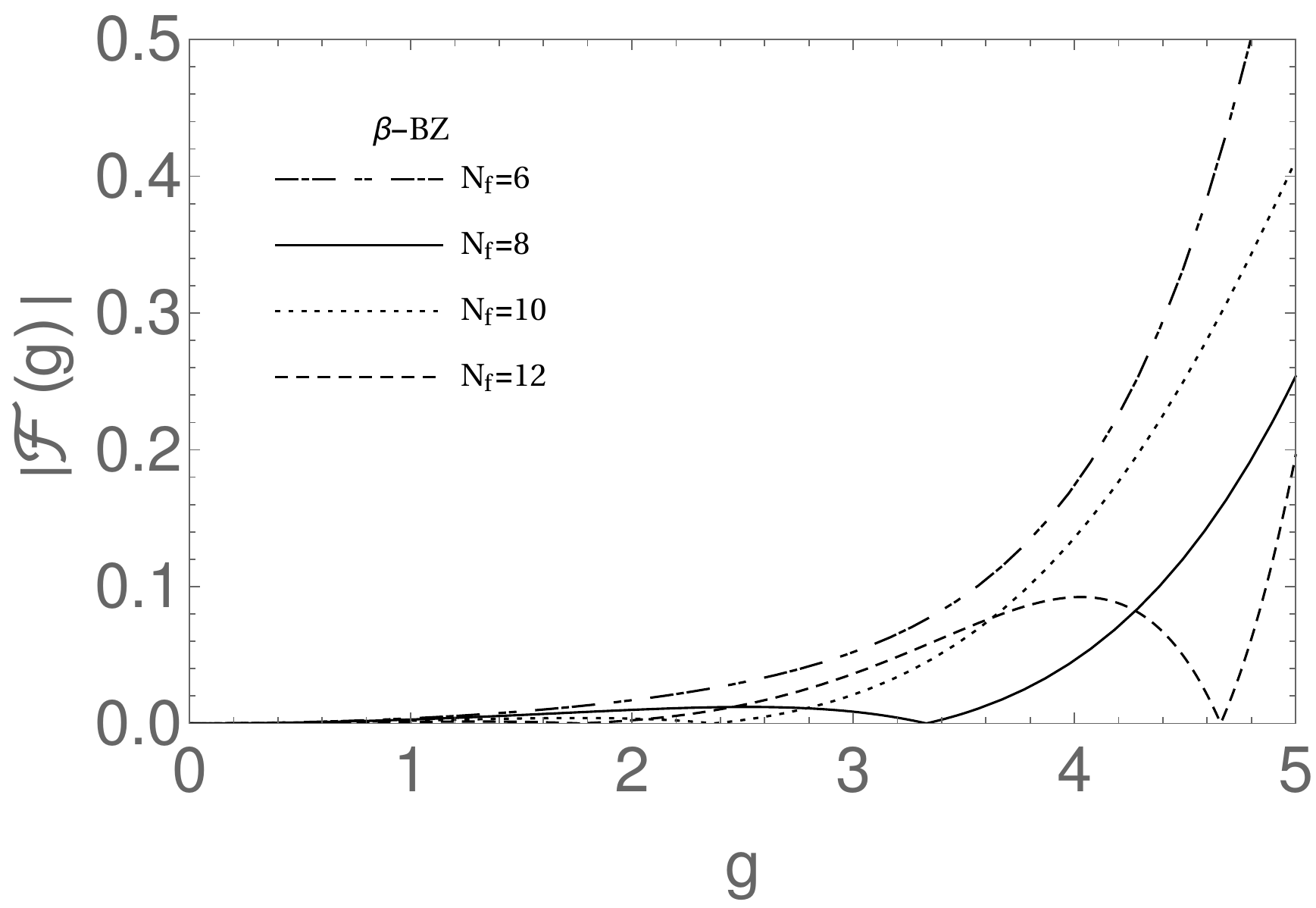}}}\label{fig:Pertu_Anali}
 \caption{Analyticity condition for both $\beta$ functions (BZ and DGM), where ${\textsl{F}}(g)$ stands for 
$\alpha_s.[d/d\alpha_s (\beta(\alpha_s)/\alpha_s)]$. Fig (a) (DGM or Eq.\eqref{eq:betaMg_k2}) and Fig.(b) 
(perturbative or Eq.\eqref{betaper}) show the analyticity condition as a function of the coupling constant $g_s$ and for different $N_f$.}
 \label{fig:Analiticity1}         
\end{figure}

We plot the left-hand side of Eq.(\ref{eq:Analyt}) in Fig.\eqref{fig:Analiticity1}. This figure allows us to see up to what value of the coupling
constant we can rely on our results. For instance, from Fig.\eqref{fig:Analiticity1}(a) we can see that the
inequality of Eq.(\ref{eq:Analyt}) is not fulfilled in the case of Eq.\eqref{eq:betaMg_k2} with $N_f=12$ for $\alpha_s \geq 1.1$. 
Since we are particularly interested in what happens at the fixed point,
in Table \eqref{tbl:IneqFP} we show the value of the left-hand side of inequality \eqref{eq:Analyt} evaluated exactly at the fixed points for
both cases: BZ and DGM. This means that we are inside, within our approximations, of the analytic region. In particular, the
DGM $\beta$ function and the respective fixed point seems to be at the border of the analytic region. Since the derivative of  Eq.(\ref{eq:betaMg_k2}) is linear in $\alpha_s$
it is easy to understand why Eq.(\ref{eq:Analyt}) is saturated at the fixed point, where the left hand side of Eq.(\ref{eq:Analyt}) is
proportional to $d\beta{\alpha_s}/d\alpha_s$.

\begin{table}[htbp]
 \caption{Left-hand side value of the inequality \eqref{eq:Analyt} evaluated at the fixed points obtained from both $\beta$ functions (BZ and DGM).}
  {\begin{tabular}{@{}ccc@{}}\toprule
   $N_f$ \hphantom{00} &  \hphantom{00}   BZ    &  \hphantom{0000}   DGM     \\ \colrule
     3  \hphantom{00}  &  \hphantom{00} $\ast$  &  \hphantom{0000}  0.9998     \\
     6  \hphantom{00}  &  \hphantom{00} $\ast$  &  \hphantom{0000}  1.0000     \\
     8  \hphantom{00}  &  \hphantom{00} 0.1053  &  \hphantom{0000}  0.9999     \\
     9  \hphantom{00}  &  \hphantom{00} 0.0583  &  \hphantom{0000}  1.0000     \\
    10  \hphantom{00}  &  \hphantom{00} 0.0340  &  \hphantom{0000}  0.9997     \\
    12  \hphantom{00}  &  \hphantom{00} 0.0112  &  \hphantom{0000}  0.9998     \\ \botrule
 \end{tabular}
 \label{tbl:IneqFP}}
\end{table}

The BZ fixed points are in the analytic region and can be certainly considered perturbative fixed points, therefore we can
compute for these points the respective anomalous dimension. Several lattice simulations have tried to compute the quark mass anomalous dimensions ($\gamma$) associated
to these fixed points, because a large anomalous dimension may solve the many problems of Technicolor (or composite) models \cite{san}.
The mass anomalous dimension exponent $\gamma$ up to ${\cal{O}}(\alpha_s^5)$ was determined in Ref.\cite{bai} in the ${\overline{MS}}$, 
and is given by
\be
\gamma = - 2 \gamma_m =  \sum_{i=0}^\infty 2 (\gamma_m)_i a_s^{i+1} \, ,
\label{dima}
\ee
where the $(\gamma_m)_i$ can be read from Ref.\cite{bai}, and in numerical form as a function of $N_f$ we have:
\br
\gamma_m  & \approx & -  a_{os} - a_{os}^2 (4.20833-0.138889 N_f)  \nonumber \\ 
& - & a_{os}^3 (19.5156-2.28412 N_f -0.0270062 N_f^2) \nonumber \\
& - & a_{os}^4 (98.9434-19.1075 N_f +0.276163 N_f^2 \nonumber \\
& + & 0.00579322 N_f^3 ) \nonumber \\ 
& - & a_{os}^5 (559.7069 -143.6864 N_f +7.4824 N_f^2 \nonumber \\
& + & 0.1083 N_f^3 - 0.000085359 N_f^4) .
\label{gamam}
\er
where $a_{os}=\alpha_s/\pi $. 

\begin{figure}[hbt]
\setlength{\epsfxsize}{0.6\hsize} \centerline{\epsfbox{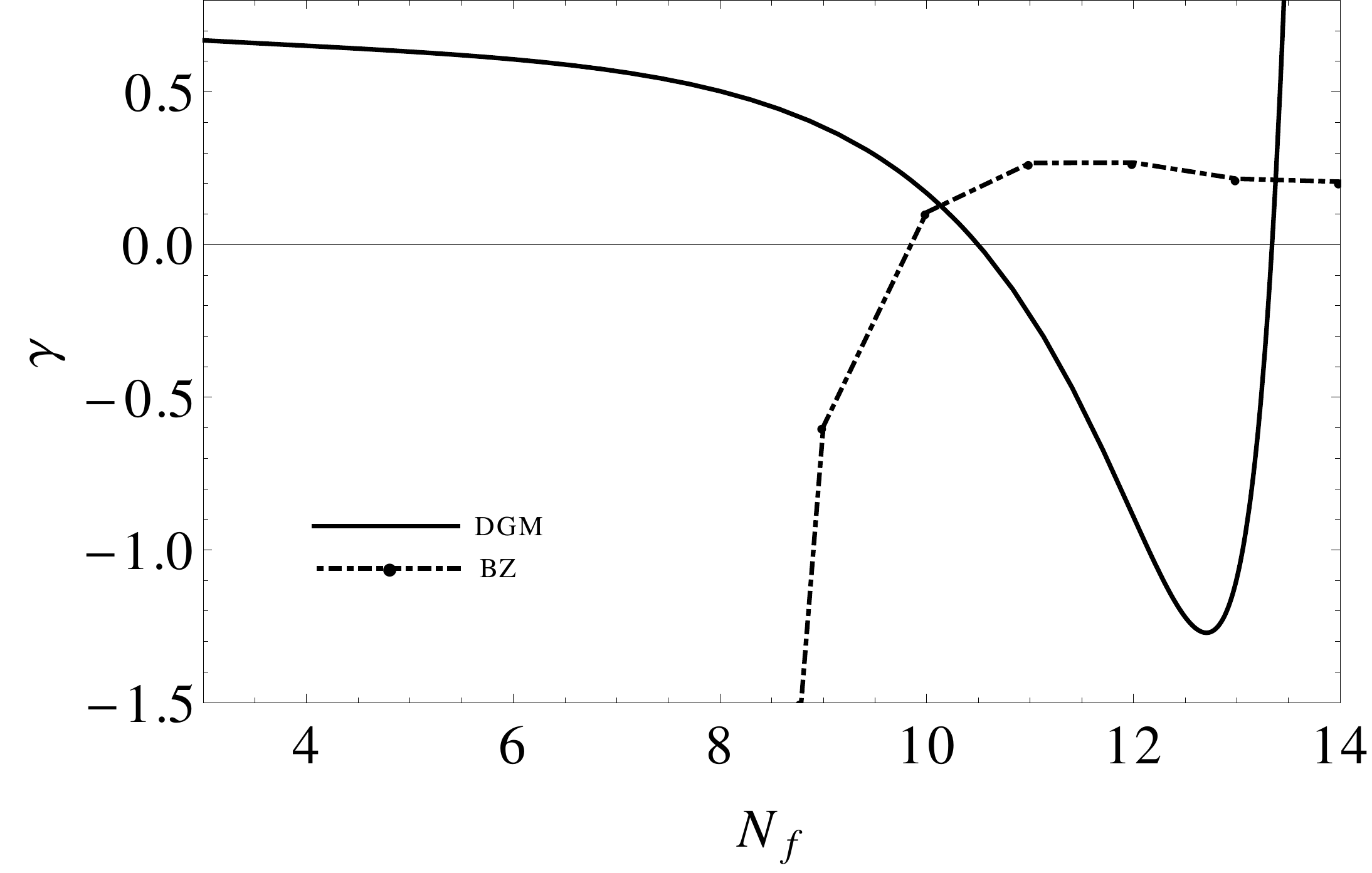}}
\caption[dummy0]{Anomalous dimension for different values of $N_f$ in the BZ and DGM cases.}
  \label{fig:GammaNf}      
\end{figure}

The DGM fixed points are at the border of the analytic region, and we may wonder if we can reliably compute the
anomalous dimension with Eq.(\ref{gamam}) even in this case. It should be remembered that chiral symmetry breaking, or the dynamical
generation of quark masses, in the presence of
dynamically generated gluon masses is still a motive of debate \cite{corc,doff}, possibly being associated to the confinement
mechanism \cite{corc}, demanding a non-perturbative anomalous dimension calculation.
However in the sequence we will just assume that we can
use Eq.(\ref{gamam}) to compute the anomalous dimensions at the DGM fixed points.
The results are shown in Fig.\eqref{fig:GammaNf}

It is interesting to see in Fig.\eqref{fig:GammaNf} that the anomalous dimension for the BZ and DGM $\beta$ functions have
different behaviors as a function of $N_f$ and this effect may probably be tested in lattice simulations.
We show in Table \eqref{tbl:gamma} the anomalous dimensions for the BZ and DGM cases respectively for
$N_f= 8, \, 9, \, 10, \, 12$.  However, note
that, if Eq.(\ref{gamam}) is applied to the DGM case, we do have large $\gamma$ values for a small number of quarks. In general it is said that such values may be present in
walking gauge theories, but this is certainly not the QCD case with a small number of quarks.  

\begin{table}[htbp]
 \caption{Anomalous dimensions evaluated at the fixed points obtained from Eq. \eqref{eq:betaMg_k2} for different values of $N_f$.  $\alpha_s^{\ast}$ is the fixed point value of the
              coupling constant for each $N_f$. }
 {\begin{tabular}{@{}lcccc@{}}
 \toprule
            &    \multicolumn{2}{c}{BZ} \hphantom{00}  & \multicolumn{2}{c}{DGM}  \\
         \cline{2-5}       
  $N_f$ \hphantom{00} & $\alpha_s^{\ast}$  \hphantom{0} &   $\gamma(\alpha_{s}^\ast)$ \hphantom{00} &   $\alpha_s^{\ast}$ \hphantom{0} &     $\gamma(\alpha_{s}^\ast)$ \\ \colrule
   8    \hphantom{00} &        1.55   \hphantom{0}      &       -4.83   \hphantom{00}               &      0.64 \hphantom{0}           &    0.501     \\
   9    \hphantom{00} &        1.07   \hphantom{0}      &       -0.60   \hphantom{00}               &      0.70 \hphantom{0}           &    0.39     \\
  10    \hphantom{00} &        0.82   \hphantom{0}      &        0.09   \hphantom{00}               &      0.78 \hphantom{0}           &    0.17     \\
  12    \hphantom{00} &        0.47   \hphantom{0}      &        0.27   \hphantom{00}               &      1.06 \hphantom{0}           &   -0.88     \\ \botrule
  \end{tabular}
  \label{tbl:gamma}}
\end{table}

In principle these anomalous dimension can be tested in lattice simulations, although they demand simulations with 
extremely large volume lattices, since the calculation must be performed in a conformal regime. There are results
for the anomalous dimension with $N_f=12$ \cite{Appel, Aoki}, indicating a value in the range $\gamma_m \approx 0.4\,-\,0.5$.  
Unfortunately this value is a factor of $2$ above the one predicted in the BZ case, and curiously is exactly the region
of the $\gamma$ values in the DGM case, although this is true only up to $N_f\approx 10$. It should also be remembered
that these lattice simulations make use of the naive hyperscaling function $M_H \propto m^{1/(1+\gamma)}$ \cite{mira} determined
for a walking gauge theory. It is possible that physical masses necessarily do not follow such scaling, and, in particular, in
the case of scalar masses we have been advocating that these masses may scale differently according to the asymptotic behavior
of the dynamically generated fermion mass \cite{and1,and2,and3}.

We end this section stressing that the BZ approach is analytic (as usually claimed), but also the DGM approach seems to be reliable up to the
border of the analytic region. Therefore we assumed that in both cases we can compute the anomalous dimension with the
perturbative expression, observing quite different behaviors for the mass anomalous dimension exponent, what could be
observed in lattice simulations.

\section{Minimum of energy: BZ or DGM?}

If the $\beta$ functions in these two approaches are comparable and lead to approximately the same fixed points for some $N_f$ values, can we determine which one leads to the actual
minimum of energy? The $\beta$ function can be related to the trace of the energy momentum tensor \cite{cre,cha,col}
\be
\left< \theta_{\mu\mu} \right> = \frac{\beta (g)}{g} \left<G_{\mu\nu}G^{\mu\nu}\right> \, ,
\label{tem}
\ee
which is proportional to the vacuum energy $\left< \Omega \right>$ as
\be
\left< \Omega \right> = \frac{1}{4} \left< \theta_{\mu\mu} \right> .
\label{ve}
\ee
The minimum of the vacuum energy is a scheme independent quantity, and, in principle, this quantity could be used to discriminate
which $\beta$ function leads to the deepest minimum of energy.  Of course the Eq.\eqref{tem} will be computed in a quite simple approximation, not free of scheme dependence, although,
we have no reason to expect great deviations of the $\beta$ functions behaviors as we discussed before. 

In Eq.(\ref{tem}) the term $\left<G_{\mu\nu}G^{\mu\nu}\right>$ 
is proportional to the gluon condensate, which is a fully non-perturbative quantity \cite{shif}. In order to calculate the vacuum energy we must know how the gluon
condensate is modified as we change the number of flavors. Unfortunately there are not, as far as we know, lattice simulations of this quantity as a function
of $N_f$. On the other hand there are discussions about the condensate value related to the infrared behavior of the gluon propagator. For instance (see Ref.\cite{zak} and
references therein), the gluon condensate is expected to be of order $c m_t^4$, where $c$ is a  constant and $m_t$ is a tachionic mass that appears in the IR gluon propagator, which
can also be related to the confining potential.

One expression for the gluon condensate as a function of the dynamical gluon mass was determined in Ref.\cite{cor1} and also studied in Ref.\cite{gor} (see Eq.(6.17) of
Ref.\cite{cor1}):
\be
\left<\frac{\alpha_s}{\pi}G_{\mu\nu}G^{\mu\nu} \right>=A\frac{3m_g^4}{4\pi^4 \beta_0 \ln\Big(\frac{4m_g^2}{\Lambda^2}\Big)},
\label{eq:gluoncondensate}
\ee
where we are going to assume that $m_g$ is a function of $N_f$, as described by the exponential behavior of Eq.(\ref{eq:MgExpPRD}) with $m_{g_0}=440$ MeV, and A ($\approx 7$)
is a constant value such that $\left<\frac{\alpha_s}{\pi}G_{\mu\nu}G^{\mu\nu} \right>=0.012$ GeV$^4$ when $N_f=0$ \cite{shif}.
This expression for the condensate is able to represent the gross behavior of this quantity as we vary $N_f$. For a fixed $N_f$ value and at one specific $g$ value  of the coupling
constant we can say that the state of minimum energy will happen at the smallest value of $\beta (g)$, although a complete answer to this problem will demand a detailed calculation
of Eq.(\ref{tem}).  Therefore, with Eqs.\eqref{tem} and introducing \eqref{eq:gluoncondensate} into Eq.\eqref{ve}, we have
\be
\left< \Omega \right> = \frac{3 A}{4\pi^2}\frac{\beta(g)}{g}m_g^4(N_f).
\label{eq:vacuum}
\ee
The assumption of Eq.(\ref{eq:gluoncondensate}), a monotonically increasing with $N_f$ function, is not fundamental to determine the minimum of energy, which is basically dictated by
the behavior of the $\beta$ function.

\begin{figure}[htb]
\centering
 \subfigure[]
           {\resizebox{0.45\columnwidth}{!}{\includegraphics[width=0.4\textwidth,angle=0]{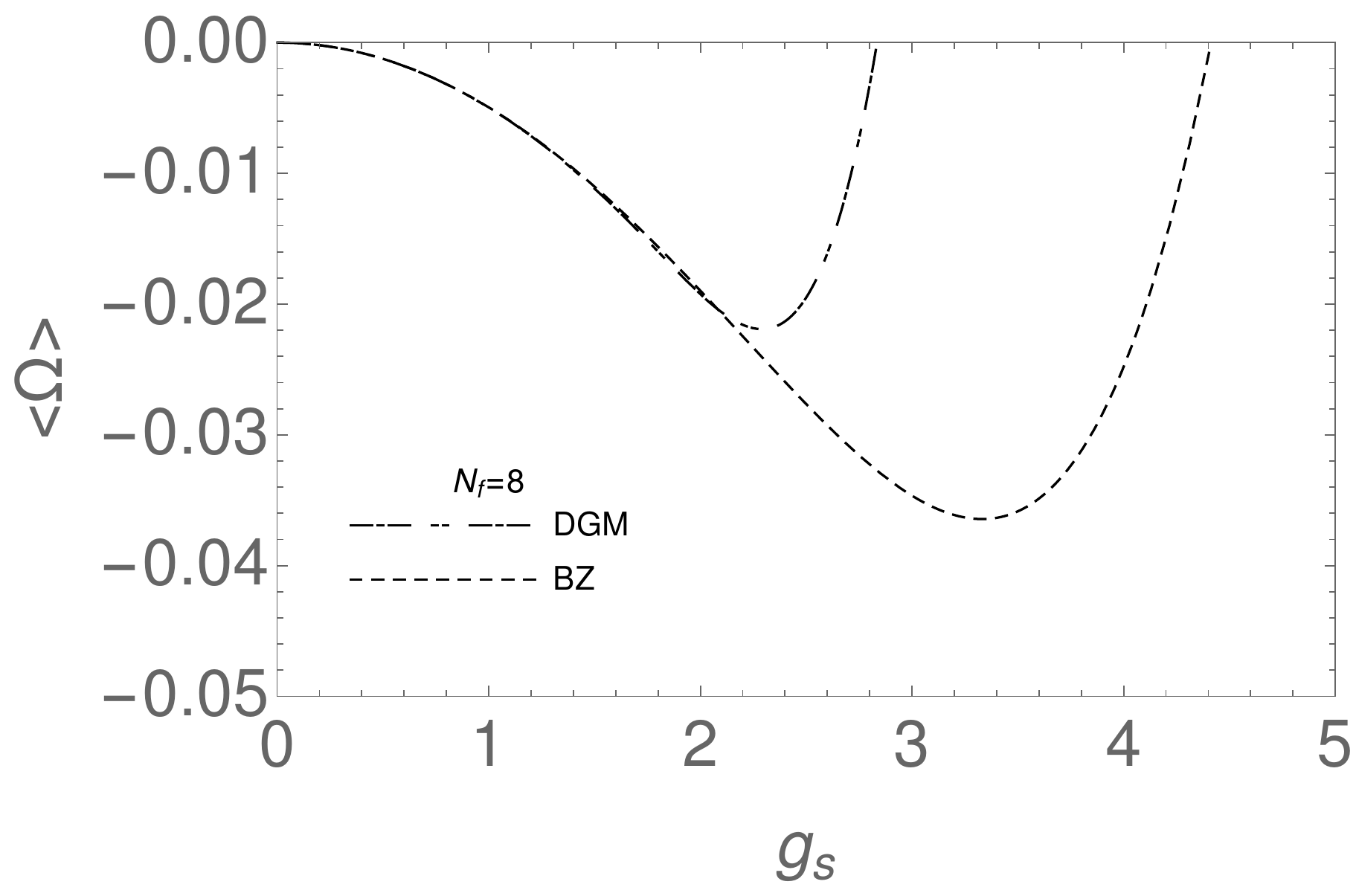}}}\label{fig:vevNf8}\\
 \subfigure[]
           {\resizebox{0.45\columnwidth}{!}{\includegraphics[width=0.4\textwidth,angle=0]{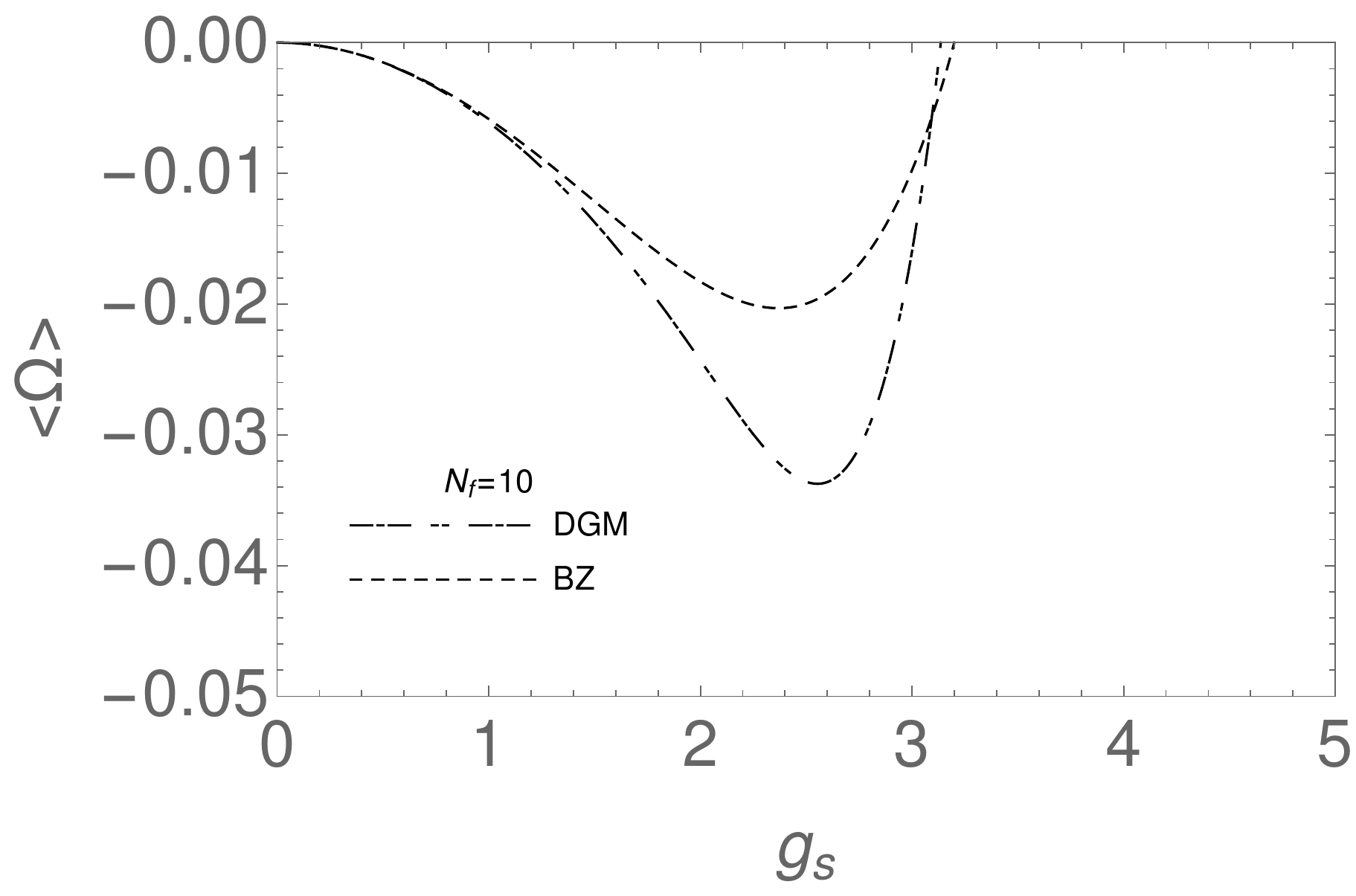}}}\label{fig:vevNf10}\\
 \subfigure[]
           {\resizebox{0.45\columnwidth}{!}{\includegraphics[width=0.4\textwidth,angle=0]{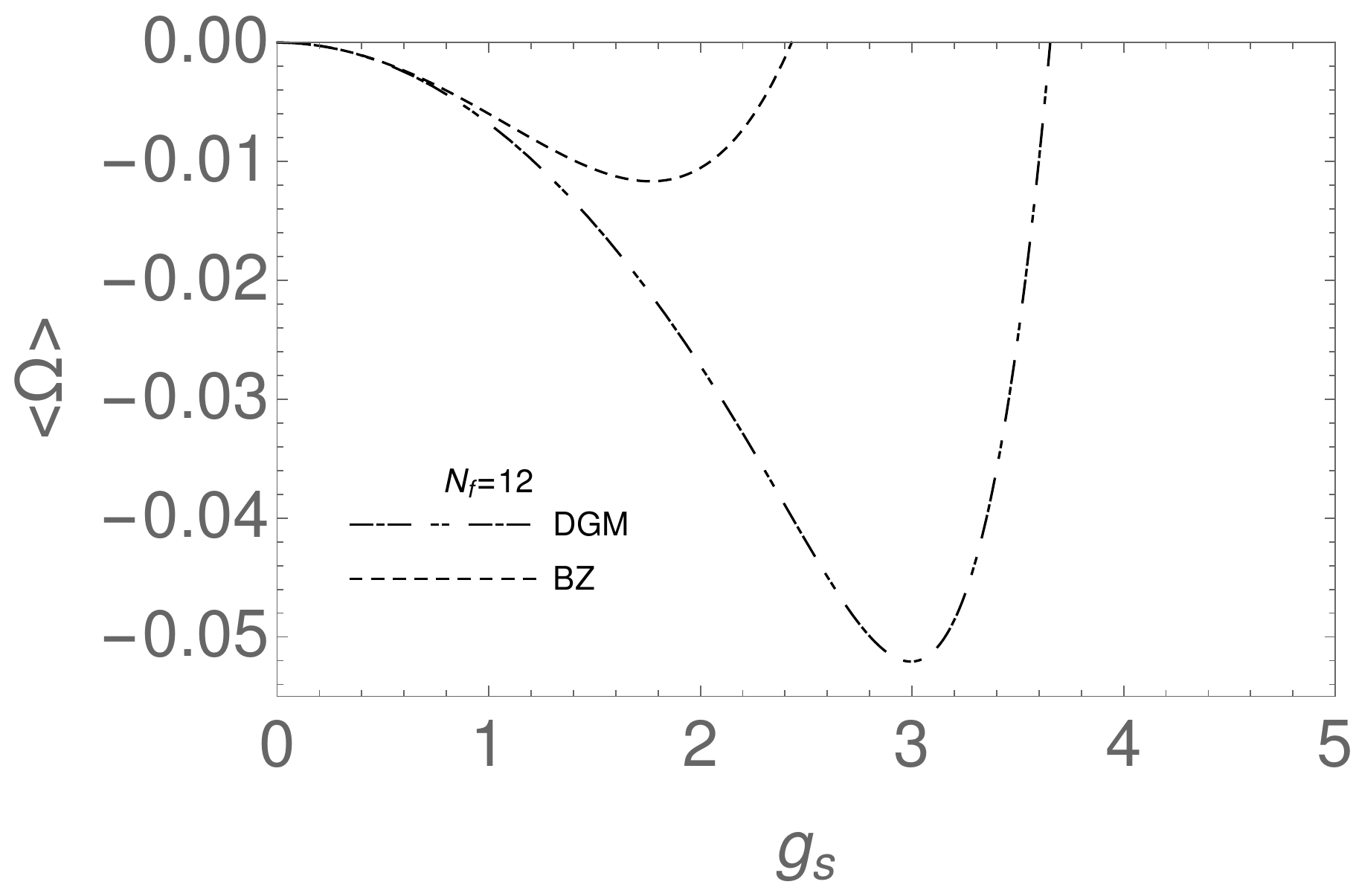}}}\label{fig:vevNf12}           
 \caption{Vacuum energy in both, BZ and DGM, approaches. Fig.(a) is for $N_f=8$, value that is in the limit of conformal window. Figs.(b) and (c) correspond to $N_f=10$
          and $N_f=12$ respectively.}
 \label{fig:veNf}         
\end{figure}

Our results for the vacuum energy as a function of the coupling constant $g_s$ are shown in Fig.(\ref{fig:veNf}) for $N_f \approx 8 \, - \, 12$. Our results are surprising in the
following sense: For $N_f=8$ it seems that the BZ $\beta$ function is the one that leads to the deepest minimum of energy as a function of the coupling constant up to the critical
$g^*$ value, although $N_f=8$ is at the border of the conformal window and below this value the ``perturbative" vacuum becomes unstable, i.e. if we define the vacuum energy
proportional to the $\beta_{BZ}$ function the vacuum is negative up to infinity for $N_f<8$. Above this $N_f$ value it is the DGM $\beta$ function that leads to the deepest state of
minimum energy. As we increase $N_f$ above $12$ it is the coupling constant in the DGM approach that increases at one point that we cannot be sure how much the SDE truncation,
leading to this solution, is still reliable. However, there is a clear possibility that the non-trivial fixed point observed in lattice simulations at $N_f=12$ is related to the DGM
mechanism.

We recall that a full calculation of the vacuum energy can be performed with the effective
potential for composite operators as a function of the complete QCD propagators \cite{cjt}. In this type of calculation besides the contribution
of dynamical gluon masses it should also be considered the inclusion of dynamical fermion masses (see, for instance, Ref.\cite{gor}).
Our simple estimate of the vacuum energy does not consider the effect of fermions, however this effect gives just a few percent contribution
to the vacuum energy for a small number of flavors \cite{gor}. As we increase the number of flavors the chiral symmetry is recovered, i.e. the dynamical generation
of fermion masses is probably erased \cite{for,tom,cap}, and the effect of fermions will probably not affect the value of
 $\left< \Omega \right>$ calculated for a number of fermions around $8-12$, which is the region where the BZ and DGM can be compared. Finally,
we also do not know how confinement may affect the dynamical masses of gluons and quarks and modify our simple estimate of the vacuum energy. 

\section{CONCLUSIONS}

We have compared two different mechanisms (BZ and DGM) proposing the existence of non-trivial fixed points in QCD. The BZ approach is essentially perturbative while the DGM
one is non-perturbative. Their $\beta$ functions are quite similar and the fixed points occur at approximately the same coupling constant values for $N_f \approx 9-10$.
However, as we vary $N_f$ the values of the coupling constants associated to the fixed points move in different directions: Decreasing when 
$N_f$ increases in the BZ
approach, and exactly the opposite occurs in the DGM case.

Both $\beta$ functions and their coupling constants up to the critical value that determine the fixed points are in agreement with the analyticity constraint, and if we assume
that in both cases we can use perturbation theory to compute the anomalous dimension associated to each fixed point, we come to the conclusion that the DGM mechanism could be a
possible explanation for the anomalous dimension and fixed points within the conformal window.

Our results cover the following number of flavors $N_f\approx 8\, - \,12$. Assuming that the gluon condensate can be calculated in terms of the non-perturbative
gluon propagator and its dynamical gluon mass, we observe an intriguing behavior of the vacuum energy calculated as a function of the
coupling constant up to the fixed point value: Around $N_f=8$ it is the BZ approach that leads to the deepest
minimum of energy, which is at the border of the conformal window, and below this number of flavors the ``perturbative" vacuum becomes unstable, in the sense that when $N_f<8$ the
BZ $\beta$ function is not bounded from below. As we increase $N_f$ the vacuum energy is dominated by the DGM mechanism. At larger $N_f$ values the coupling constant in the DGM
approach seems to increase to values where we cannot be sure that the SDE calculations
of the DGM approach are still reliable. Of course, in the DGM case we do have fixed points for a small number of quarks, and for
a naive (perturbative) calculation of the mass anomalous dimension exponent we obtain $\gamma$ in the range $0.4-0.5$ up to $N_f \approx 10$.
If the fixed points predicted in the DGM approach are not confirmed by lattice simulations this means that the extrapolations shown in Eqs. \eqref{eq:MgLinearPRD} and
\eqref{eq:MgExpPRD} are not correct or the $\beta$ function of Eq.\eqref{eq:betacornwall} does not correspond to a true minimum of energy.

It should be noted that we used simple approximations for the coupling constant (Eq.(\ref{eq:Coupling})), for the dynamical gluon mass behavior with the momentum (Eq.(\ref{eq:mg})), 
for the dependence of this mass (Eq.(\ref{eq:MgExpPRD})) and of the gluon condensate (Eq.(\ref{eq:gluoncondensate})) with the number of flavors. The dependence of these quantities
with $N_f$ introduces some uncertainty in our calculation, which are difficult to be estimated, but shall not modify the main results and characteristics of the DGM mechanism.
In particular, it has been strongly stressed how lattice simulations  of the gluon propagator, and consequently the dynamical gluon mass IR value, demand simulations with quite large
volume lattices \cite{cuca}. We also note that our calculation depends on the ratio $m_g/\Lambda$, and the $\Lambda$ value will be dependent on the scheme and the number of fermions
that we consider in order to obtain its physical value. If we consider recently reported $\Lambda^{(N_f)}_{\bar{MS}}$ determinations obtained in lattice simulations (e.g. 
$\Lambda^{(2)}_{\overline{MS}} = 330^{+21}_{-54}$ MeV and $\Lambda^{(3)}_{\overline{MS}} = 336 (19)$ MeV \cite{Aoki1}), we can be confident that the uncertainty that we have in the $m_g$
determination exceed by far the one that we have for $\Lambda$, and for this reason we stress that our result is quite dependent on the ratio $m_g/\Lambda$ and the main source of
uncertainty resides in the $m_g$ variation with $N_f$.
We are comparing $\beta$ functions obtained in different schemes, and are assuming that their respective fixed points are relatively stable, making our
comparison worthwhile.  Considering the proximity of the different fixed points, and their dependence with $N_f$ it would be important that such coincidence could be tested by
different methods.  Therefore, it is imperative to have improved lattice calculations of the behavior of this quantity with large $N_f$, mainly to check extrapolations like the ones
of Eq.\eqref{eq:MgLinearPRD} and  Eq.\eqref{eq:MgExpPRD}.  The same can also be said about calculations of the gluon condensate as a function of $N_f$. This will allow better
estimates of the fixed points as well as of the vacuum energy.

\section*{Acknowledgments}
\vspace{-0.5cm}
This research was partially supported
by the Conselho Nac. de Desenv. Cient\'{\i}fico e Tecnol\'ogico
(CNPq), by the grants 2013/22079-8 and 2013/24065-4 of Funda\c c\~ao de Amparo \`a Pesquisa do
Estado de S\~ao Paulo (FA\-PES\-P) and by Coordena\c c\~ao de Aper\-fei\-\c coa\-mento
de Pessoal de N\'{\i}vel Superior (CAPES).

\end{document}